\documentclass[12pt,aps,superscriptaddress,nofootinbib,notitlepage]{revtex4-1}
\usepackage[dvips]{graphicx}
\usepackage{latexsym,amssymb,amsmath,bm}
\usepackage{color}
\usepackage[bookmarksnumbered,bookmarksopen,colorlinks,citecolor=red,linkcolor=blue]{hyperref}
\usepackage{mathrsfs}
\usepackage{times}
\usepackage{csquotes}

\renewcommand\a{\alpha}
\renewcommand\b{\beta}
\renewcommand\d{\delta}

\renewcommand\l{\lambda}
\renewcommand\r{\rho}

\renewcommand\t{\tau}

\renewcommand\c{\chi}
\renewcommand\j{\psi}
\renewcommand\o{\omega}
\newcommand\e{\epsilon}
\newcommand\g{\gamma}
\newcommand\z{\zeta}
\newcommand\m{\mu}
\newcommand\n{\nu}
\newcommand\x{\xi}
\newcommand\p{\pi}
\newcommand\h{\theta}
\newcommand\s{\sigma}
\newcommand\f{\phi}
\newcommand\w{\eta}
\newcommand\ve{\varepsilon}

\renewcommand\S{\Sigma}

\renewcommand\H{\Theta}
\newcommand\D{\Delta}
\newcommand\G{\Gamma}
\newcommand\F{\Phi}

\newcommand{\eq}[1]{Eq.~(\ref{#1})}

\newcommand{\eqs}[2]{Eqs.~(\ref{#1})-(\ref{#2})}

\newcommand\lb{\left(}
\newcommand\rb{\right)}
\newcommand\ls{\left[}
\newcommand\rs{\right]}
\newcommand\lc{\left\{}
\newcommand\rc{\right\}}
\newcommand{\lan}{\langle}
\newcommand{\ran}{\rangle}

\newcommand\ra{\rightarrow}

\newcommand{\non}{\nonumber\\}
\newcommand\pt{\partial}

\newcommand{\cl}{{\cal L}}

\newcommand{\Tr}{{\rm Tr}}

\newcommand{\bp}{{\vec p}}
\newcommand{\bk}{{\vec k}}
\newcommand{\bq}{{\vec q}}

\renewcommand{\part}{{\rm part}}
\renewcommand{\vec}[1]{{\bm #1}}

\begin{document}

\title{An introduction to relativistic spin hydrodynamics}

\author{Xu-Guang Huang}
\email{huangxuguang@fudan.edu.cn}
\affiliation{Physics Department and Center for Particle Physics and Field Theory, Fudan University, Shanghai 200438, China}
\affiliation{Key Laboratory of Nuclear Physics and Ion-beam Application (MOE), Fudan University, Shanghai 200433, China}
\affiliation{Shanghai Research Center for Theoretical Nuclear Physics, National Natural Science Foundation of China and Fudan University, Shanghai 200438, China}
\date{\today}

\begin{abstract}
Spin polarization and spin transport are common phenomena in many quantum systems. Relativistic spin hydrodynamics provides an effective low-energy framework to describe these processes in quantum many-body systems. The fundamental symmetry underlying relativistic spin hydrodynamics is angular momentum conservation, which naturally leads to inter-conversion between spin and orbital angular momenta. This inter-conversion is a key feature of relativistic spin hydrodynamics, closely related to entropy production and introducing ambiguity in the construction of constitutive relations. In this article, we present a pedagogical introduction to relativistic spin hydrodynamics. We demonstrate how to derive the constitutive relations by applying local thermodynamic laws and explore several distinctive aspects of spin hydrodynamics. These include the pseudo-gauge ambiguity, the behavior of the system in the presence of strong vorticity, and the challenges of modeling the freeze-out of spin in heavy-ion collisions. We also outline some future prospects for spin hydrodynamics.
\end{abstract}
\maketitle

\section{Introduction}\label{sec:intro}

Spin is a fundamental property of particles arising from quantum mechanics, playing a central role in numerous phenomena within the quantum regime. As a form of angular momentum, spin naturally couples to rotation, allowing it to become polarized by rotational motion. Similarly, for a charged particle with nonzero spin, or a neutral particle with a non-trivial charge form factor, spin can couple to an external magnetic field as well. Additionally, for a particle in motion (i.e., with finite momentum), its spin may couple to acceleration, electric fields, or gradients of external potentials, such as a chemical potential and temperature. In the case of massless particles, the spin state is specified by its helicity state, meaning it is intrinsically slaved to the particle's motion. As a result, spin can be manipulated by rotating fields, magnetic fields, electric fields, and a number of other external influences. Conversely, detecting the spin of a particle provides invaluable insights into the environment or the underlying dynamics of the system.

In heavy-ion collision physics, the primary interest lies in the creation of deconfined quark-gluon matter, commonly referred to as the quark-gluon plasma (QGP)~\cite{Wang:2016opj,DU2023Review,ZHANG2023Experimental,Shou:2024uga,Chen:2024aom}. To uncover the properties of QGP in heavy-ion collision experiments, it is essential to design specific hadronic observables that are sensitive to particular features of the QGP. Since charged particles are typically the easiest to detect, many observables rely on the charge of the hadrons. For instance, the total multiplicity of detected charged hadrons reflects the initial energy of the QGP. Meanwhile, the anisotropy in the momentum-space distribution of charged hadrons corresponds to the initial anisotropy in the spatial distribution of partons, leading to the well-known harmonic flow parameters~\cite{Ollitrault:1992bk}. By measuring these hadronic observables, people have revealed several novel properties of the hot and dense matter created in heavy-ion collisions. One significant finding is that the QGP must be extremely hot, with a typical temperature reaching 300-500 MeV at RHIC and LHC, indicating an extremely large energy density. Additionally, the QGP medium is found to be strongly interacting, with a very small shear viscosity to entropy density ratio $\eta/s$. This low ratio is required to explain the observed harmonic flow parameters~\cite{Shou:2024uga,Chen:2024aom}. In fact, the $\eta/s$ of QGP is the lowest among all known fluids.

Since 2017, it has been established that the spin degree of freedom can also be used to probe the properties of the QGP~\cite{STAR:2017ckg}. This is achieved through measurements of the spin polarization of spinful hadrons, such as hyperons and vector mesons~\cite{Liang:2004ph,Voloshin:2004ha,Liang:2004xn}. Notably, it has been observed that the $\Lambda$ and $\bar{\Lambda}$ hyperons can exhibit significant spin polarization at collision energies of tens of GeV~\cite{STAR:2017ckg,STAR:2018gyt,STAR:2021beb,HADES:2022enx,STAR:2023nvo}. Similarly, the $\phi$ and $J/\psi$ mesons have been found to exhibit considerable spin alignment~\cite{STAR:2022fan,ALICE:2022dyy}~\footnote{The spin alignment of a vector meson is quantified by the deviation of $\rho_{00}$ from $1/3$, where $\rho_{00}$ is the $00$-component of the vector meson's spin density matrix.}. These discoveries open new avenues for studying QGP through the spin degree of freedom. For instance, we now understand that the so-called global spin polarization (i.e., the total amount of spin polarization with respect to the reaction plane) of hyperons arises from angular momentum conservation through the formation of fluid vortices within the QGP: In non-central heavy-ion collisions, the system possesses substantial orbital angular momentum, which subsequently induces strong fluid vorticity in the QGP~\cite{Deng:2016gyh,Jiang:2016woz,Deng:2020ygd}, thereby polarizing the spins of quarks via spin-rotation coupling~\cite{Becattini:2013fla,Fang:2016vpj,Liu:2020flb,Karpenko:2016jyx,Xie:2017upb,Li:2017slc,Shi:2017wpk,Xia:2018tes,Wei:2018zfb,Vitiuk:2019rfv,Ivanov:2019ern,Fu:2020oxj,Guo:2021uqc,Li:2021zwq,Deng:2021miw,Wu:2022mkr}. However, to fully understand the spin polarization phenomena, a dynamical theory of spin polarization and spin transport in the hot medium is essential, analogous to the necessity of a dynamical theory of the bulk medium for understanding harmonic flows. Naturally, such a dynamical theory of spin transport could be derived from either kinetic theory or hydrodynamic theory. In recent years, both spin kinetic theory and spin hydrodynamics have made significant advancements. In this article, we will focus on spin hydrodynamics, while we refer readers to Refs.~\cite{Hidaka:2022dmn} for a review of spin kinetic theory and to Refs.~\cite{Gao:2020vbh,Becattini:2020ngo,Huang:2020dtn,Huang:2020xyr,Liu:2020ymh,Becattini:2022zvf,Xin-Li:2023gwh,Jian-Hua:2023cna,Li-Juan:2023bws,Jian-Hua:2023cna,Jian-Hua:2023viv,Becattini:2024uha,Chen:2024afy,Xu:2023lnd} for a review of spin polarization phenomena in heavy-ion collisions. Besides, we will focus only on spin polarization in hot and dense medium rather than in systems created in, e.g., electron-ion collisions~\cite{Ji:2023cdh}.

Throughout this article, we use the natural units $c=\hbar=k_B=1$ and the metric convention $\eta_{\m\n}=\eta^{\m\n}={\rm diag}(1,-1,-1,-1)$.

\section{Relativistic hydrodynamics as an effective theory}\label{sec:hydro}

Before we go into the discussion of spin hydrodynamics, let us first briefly review the general structure of relativistic hydrodynamics from the perspective of effective field theory. Hydrodynamic theory describes the low-energy behavior of interacting many-body systems, where only conserved charge densities exhibit their dynamics. Since conserved charge densities do not vanish, they simply redistribute themselves in space according to their equations of motion (EOMs).  When expressed in a manner of spatial gradient expansion, these EOMs constitute the hydrodynamic equations.

Let us consider the hydrodynamic theory of a system with spacetime translation symmetry and a global $U(1)$ symmetry. The corresponding conserved charge densities are the energy density $\ve(x)$, momentum density $\p^i(x), i=1-3$, and the $U(1)$ charge density $n(x)$. We want to derive the dynamical equations for these conserved charge densities. Sometimes it is more convenient to work with the potential variables conjugate to the charge densities. They are the temperature $T(x)$ (or its inverse $\b(x)=1/T(x)$), the fluid velocity $u^\m(x)$ normalized as $u^\m u^\n \eta_{\m\n}=1$, and the chemical potential of $n(x)$, $\m(x)$. These conserved charge densities (and equivalently their conjugates) are hydrodynamic variables in hydrodynamics. Our starting point is the conservation laws:
\begin{eqnarray}
\label{con:tmn}
\pt_\m \H^{\m\n}=0,\\
\label{con:j}
\pt_\m J^\m=0,
\end{eqnarray}
where $\H^{\m\n}$ is the energy-momentum tensor and $J^\m$ is the $U(1)$ current. As an effective field theory, we want to express $\H^{\m\n}$ and $J^\m$ in terms of the conserved charged density (or equivalently, their conjugates) and their various gradient orders. We assume spatial isotropy of the system, i.e., there is no external forces breaking the $SO(3)$ symmetry. The building blocks are the fluid velocity $u^\m$ and various quantities that can be classified into different representations of $SO(3)$ in the rest frame of the fluid. Up to first order in gradients, these quantities are: 
\begin{eqnarray}
\label{hydro:bb}
&&{\rm Scalar:}\quad\ve,\; n, \;D\ve, Dn,  \h\equiv\nabla\cdot u =\pt\cdot u,\; \non
&& {\rm Vector:}\quad Du^\m,\; \nabla^\m\ve, \nabla^\m n,\;\o^{\m\n}\equiv -(1/2)(\nabla^\m u^\n -\nabla^\n u^\m),\;\non
&&{\rm Tensor:}\quad \s^{\m\n}\equiv (1/2)[\nabla^\m u^\n +\nabla^\n u^\m -(2/3)\D^{\m\n}\h],\;
\end{eqnarray}
where $D\equiv u\cdot\pt$ is the co-moving time derivative, $\h$ is the expansion rate of the fluid, $\nabla_\m=\D_{\m\n}\pt^\n$ is the spatial gradient operator with $\D_{\m\n}\equiv \eta_{\m\n}-u_\m u_\n$ the spatial projector, $\s^{\m\n}$ is the shear tensor which is traceless, and $\o_{\m\n}$ is the vorticity tensor. Note that the co-moving time derivatives will be eventually replaced by spatial gradients up on using the EOMs at leading order. Note that the vorticity tensor transform the same way as a three-vector under proper three-rotations (i.e., a three-rotation $R$ with $\det R=1$) as it can be substituted by a three pseudo-vector $\o^\m\equiv -(1/2)\e^{\m\n\r\s}u_\n\o_{\r\s}$. Consequently, we can write down the most general structure decomposition up to $O(\pt)$ for $\H^{\m\n}$ and $J^\m$ as follows~\footnote{One can start without including the co-moving time-derivative terms as those terms are eventually replaced by the spatial gradients up on using leading-order hydrodynamic EOMs. But we keep them to make the discussions more transparent.}:
\begin{eqnarray}
\label{tmn:exp}
\H^{\m\n} &=& (a_0+b_0^\ve D\ve +b^n_0 D n +b^u_0\h) u^\m u^\n\non
&&+c_0 (u^\m\nabla^\n\ve+u^\n\nabla^\m\ve) + d_0 (u^\m\nabla^\n n +u^\n\nabla^\m n) +e_0 (u^\m D u^\n+u^\n D u^\m) + f_0(u^\m\o^\n+u^\n\o^\m)\non
&&+(g_0+h^\ve_0 D\ve +h^n_0 D n +h^u_0\h) \D^{\m\n} \non
&& + i_0 \s^{\m\n}\non
&&+j_0 (u^\m\nabla^\n\ve-u^\n\nabla^\m\ve) + k_0 (u^\m\nabla^\n n -u^\n\nabla^\m n) +l_0 (u^\m D u^\n-u^\n D u^\m) + m_0(u^\m\o^\n-u^\n\o^\m)\non
&&+n_0 \e^{\m\n\r\s}u_\r\nabla_\s\ve +o_0 \e^{\m\n\r\s}u_\r\nabla_\s n+p_0 \e^{\m\n\r\s}u_\r D u_\s+q_0\o^{\m\n}\non &&+ O(\pt^2),\\
\label{j:exp}
J^\m &=& (A_0+B_0^\ve D\ve +B^n_0 D n +B^u_0\h)u^\m + C_0 \nabla^\m \ve + D_0\nabla^\m n + E_0 D u^\m + F_0\o^\m + O(\pt^2),
\end{eqnarray}
with all the coefficients (playing roles of the Wilson coefficients in effective field theory, as the short-distance physics are encoded in these coefficients) functions of $\ve$ and $n$. They are constructed by  decomposing first with respect to $u^\m$ and then with respect to different representation of $SO(3)$. In these decompositions, the terms with $f_0, m_0, n_0, o_0, p_0$ in $\H^{\m\n}$ and $F_0$ in $J^\m$ as coefficients transform differently from $\H^{\m\n}$ and $J^\m$ under parity (P), respectively, meaning that they can appear only when the system contains parity violating contents. Under time reversal transformation (T), all the terms of first-order gradients on the right-hand sides of $\H^{\m\n}$ and $J^\m$ except for terms with coefficients $f_0, m_0, n_0, o_0, p_0, F_0$ transform differently from $\H^{\m\n}$ and $J^\m$, respectively. This means that these terms must be dissipative (that is to say, these terms are responsible for entropy generation in the fluid), while terms with coefficients $f_0, m_0, n_0, o_0, p_0, F_0$ can appear without generating entropy, i.e., they could arise in ideal hydrodynamics despite they are at first order in gradients. The terms with coefficients $f_0, m_0, n_0, o_0, p_0, F_0$ are thus especially interesting. In fact, some of them have been intensively studied and it was found that they contain very rich quantum phenomena (usually dubbed chiral anomalous transports) closely related to chiral anomaly of the system if the underlying physics is governed by quantum gauge theory. Recently, such chiral anomalous transports have been an active subjects in condensed matter physics, astrophysics, and heavy-ion collision physics; see Refs.~\cite{Kharzeev:2015znc,Huang:2015oca,Hattori:2016emy,Liu:2020ymh,Kharzeev:2020jxw,Xin-Li:2023hqa,Kharzeev:2024zzm} for recently reviews with a focus on heavy-ion collision physics. Similarly, we could also examine the balance of right-hand and left-hand sides of \eqs{tmn:exp}{j:exp} under charge conjugation (C) transformation. The terms with coefficients $b_0^n, d_0, h_0^n, k_0, o_0, A_0, B_0^\ve, B_0^u, C_0, E_0, F_0$ must vanish if there is no environmental charge-conjugation violation (Naturally, the presence of a nonzero charge density $n$ violates the C symmetry and allows these terms to present). The antisymmetric terms in $\H^{\m\n}$ are also particularly interesting. To reveal their meaning, we consider the angular momentum conservation law:
\begin{eqnarray}
\label{con:Mmnr}
\pt_\m M^{\m\n\r}=0,
\end{eqnarray}
where $M^{\m\n\r}$ is the angular momentum tensor 
\begin{eqnarray}
\label{Mmnr:dec}
M^{\m\n\r}= x^\n\H^{\m\r}-x^\r\H^{\m\n}+\S^{\m\n\r},
\end{eqnarray}
with $\S^{\m\n\r}$ the spin tensor. We can re-write \eq{con:Mmnr} in the following form:
\begin{eqnarray}
\label{con:Mmnr2}
\pt_\m \S^{\m\n\r}=\H^{\r\n}-\H^{\n\r}.
\end{eqnarray}
Thus the antisymmetric part of $\H^{\m\n}$ provide a source for the generation of spin (one may more clearly see this by integrating \eq{con:Mmnr2} over space). This will be the focus of this article and we will come back to it from the next section. In the rest of this section, for the purpose of demonstrating the construction of the hydrodynamic theory, we will simply assume that the system does not possess a spin tenor, so that $\H^{\m\n}$ is symmetric, $\H^{\m\n}=\H^{\n\m}$, and will assume that there is no environmental parity violation, so that terms with coefficients $f_0, m_0, n_0, o_0, p_0, F_0$ must vanish. Thus, the most general decomposition of $\H^{\m\n}$ and $J^\m$ up to the first order in gradients into different components with respect to $u^\m$, and subsequently, for the components orthogonal to $u^\m$, with respect to different irreducible tensor structures under $SO(3)$ are as follows:
\begin{eqnarray}
\label{tmn:exp2}
\H^{\m\n} &=& (a_0+b_0^\ve D\ve +b^n_0 D n +b^u_0\h) u^\m u^\n\non
&&+c_0 (u^\m\nabla^\n\ve+u^\n\nabla^\m\ve) + d_0 (u^\m\nabla^\n n +u^\n\nabla^\m n) +e_0 (u^\m D u^\n+u^\n D u^\m) \non
&&+(g_0+h^\ve_0 D\ve +h^n_0 D n +h^u_0\h) \D^{\m\n} \non
&& + i_0 \s^{\m\n}\non
&&+ O(\pt^2),\\
\label{j:exp2}
J^\m &=& (A_0+B_0^\ve D\ve +B^n_0 D n +B^u_0\h)u^\m + C_0 \nabla^\m \ve + D_0\nabla^\m n + E_0 D u^\m  + O(\pt^2).
\end{eqnarray}

Up to this point, the expressions (\ref{tmn:exp2})-(\ref{j:exp2}) are merely a parametrization of $\H^{\m\n}$ and $J^\m$, and such a parametrization is ambiguous  at $O(\pt)$ order (and higher orders in gradients). To see this, we consider to re-express $\H^{\m\n}$ and $J^\m$ in terms of a redefinition of the hydrodynamic variables $\ve', n', u'^\m$ which differ from $\ve, n, u^\m$ by $O(\pt)$-order shifts:
\begin{eqnarray}
\label{shift:hydro}
\ve'=\ve + \d\ve,\quad n'= n + \d n,\quad u'^\m= u^\m + \d u^\m, 
\end{eqnarray}
where $\d\ve, \d n, \d u^\m$ are order-$O(\pt)$ quantities and $u_\m\d u^\m=O(\pt^2)$ so that $u'^2=1$ is kept at $O(\pt)$. This can be seen by noticing that $u'^2=u^2+2u_\m \d u^\m + \d u^2$ which leads to $2u_\m \d u^\m + \d u^2=O(\pt^2)$ and therefore $u_\m\d u^\m$ must be $O(\pt^2)$. In terms of the primed variables, we have 
\begin{eqnarray}
\label{tmn:exp3}
\H^{\m\n} &=& \ls a'_0-\lb\frac{\pt a_0}{\pt \ve}\d \ve+\frac{\pt a_0}{\pt n}\d n\rb + b_0^\ve D\ve +b^n_0 D n +b^u_0\h\rs u'^\m u'^\n\non
&&+c_0 (u^\m\nabla^\n\ve+u^\n\nabla^\m\ve) + d_0 (u^\m\nabla^\n n +u^\n\nabla^\m n) +e_0 (u^\m D u^\n+u^\n D u^\m) +(g_0-a_0) (\d u^\m u^\n +u^\m \d u^\n)\non
&&+\ls g'_0-\lb\frac{\pt g_0}{\pt \ve}\d \ve+\frac{\pt g_0}{\pt n}\d n\rb+h^\ve_0 D\ve +h^n_0 D n +h^u_0\h\rs \D'^{\m\n} \non
&& + i_0 \s^{\m\n}+ O(\pt^2),\\
\label{j:exp3}
J^\m &=& \ls A'_0-\lb\frac{\pt A_0}{\pt \ve}\d \ve+\frac{\pt A_0}{\pt n}\d n\rb+B_0^\ve D\ve +B^n_0 D n +B^u_0\h\rs u'^\m \non &&-A_0 \d u^\m+ C_0 \nabla^\m \ve + D_0\nabla^\m n + E_0 D u^\m 
\non&& + O(\pt^2),
\end{eqnarray}
where $a_0'=a_0(\ve',n')$ and similarly for $g'_0, A'_0$ and all the second order terms are omitted. By observing the expressions in the three square brackets, one can realize that, by suitably choosing $\d\ve$ and $\d n$, one can eliminate the first-order terms in two of the three square brackets. For example, one can solve out $\d\ve$ and $\d n$ by requiring the first-order terms in the square brackets in $\H^{\m\n}$ to vanish. But it is much more convenient to eliminate the first-order terms in the coefficients of $u'^\m u'^\n$ in $\H^{\m\n}$ and $u'^\m$ in $J^\m$. Similarly, by suitably choosing $\d u^\m$, one can eliminate either the second line in $\H^{\m\n}$ (such a choice is called Landau-Lifshitz frame for $u^\m$) or the second line in $J^\m$ (such a choice is called Eckart frame for $u^\m$). Therefore, we could always choose the following simpler forms for $\H^{\m\n}$ and $J^\m$ (Landau-Lifshitz frame):
\begin{eqnarray}
\label{tmn:exp4}
\H^{\m\n} &=& a_0 u^\m u^\n +(g_0+h^\ve_0 D\ve +h^n_0 D n +h^u_0\h) \D^{\m\n} + i_0 \s^{\m\n}+ O(\pt^2),\\
\label{j:exp4}
J^\m &=& A_0 u^\m + C_0 \nabla^\m \ve + D_0\nabla^\m n + E_0 D u^\m  + O(\pt^2).
\end{eqnarray}
Contracting with $u^\m$, we can identify that $a_0=u_\m u_\n\H^{\m\n}$ which is the local energy density $\ve$ and $A_0=u\cdot J$ which is the local $U(1)$ charge density $n$. [Sometimes, this is also considered as the matching condition because this means that $u_\m u_\n\H^{\m\n}=u_\m u_\n\H^{\m\n}_{(0)}$ and $u_\m J^\m=u_\m J^\m_{(0)}$ with $\H^{\m\n}_{(0)}$ and $J^\m_{(0)}$ the zeroth order energy-momentum tensor and charge current.]

Let us first consider the zeroth order terms which, as we have already discussed, correspond to ideal hydrodynamics:
\begin{eqnarray}
\label{tmn:expid}
\H^{\m\n}_{(0)} &=& \ve u^\m u^\n + g_0 \D^{\m\n} ,\\
\label{j:expid}
J^\m_{(0)} &=& n u^\m .
\end{eqnarray}
In the rest frame of the fluid, $u^\m=(1,\bm 0)$, it becomes $\H^{\m\n}_{(0)}={\rm diag}(\ve,-g_0,-g_0,-g_0)$ which identifies $-g_0$ to be the thermodynamic pressure $P$. At zeroth order, the conservation laws read
\begin{eqnarray}
\label{zero:euler1}
&(\ve+P)D u^\m-\nabla^\m P=0, &\\
\label{zero:euler2}
&D\ve +(\ve+P)\h=0,&\\
\label{zero:n}
&D n+n\h=0.&
\end{eqnarray}
To close these equations, we need to know the thermodynamic relation among $P,\ve, n$, that is, the equation of state, $P=P(\ve, n)$.

Let us then consider the first-order terms which correspond to dissipative hydrodynamics. From \eqs{zero:euler1}{zero:n}, we notice that we could replace $D\ve$ and $Dn$ in the first order terms by $-(\ve+P)\h$ and $-n\h$ and $Du^\m$ by $\nabla^\m P/(\ve+P)$. This allows us to re-write energy-momentum tensor and charge current as:
\begin{eqnarray}
\label{tmn:exp5}
\H^{\m\n} &=& \ve u^\m u^\n -(P+h_0\h) \D^{\m\n} + i_0 \s^{\m\n}+ O(\pt^2),\\
\label{j:exp5}
J^\m &=& n u^\m + C'_0 \nabla^\m \ve + D'_0\nabla^\m n + O(\pt^2),
\end{eqnarray}
with $h_0=(\ve+P)h_0^\ve+n h_0^n-h_0^u$, $C'_0=C_0+E_0\lb\pt P/\pt\ve\rb_n/(\ve+P)$, and $D_0'=D_0+E_0\lb\pt P/\pt n\rb_\ve/(\ve+P)$. Further constraints can be imposed from the laws of local thermodynamics. For a fluid at rest, we have the first law of thermodynamics as
\begin{eqnarray}
\label{ther1}
& Tds +\m dn= d\ve,&\\
&Ts+\m n=\ve+P,&
\end{eqnarray}
with $s$ the entropy density. To proceed, we propose the covariant generalization of the second one (the Gibbs-Duhem relation):
\begin{eqnarray}
\label{gbrelation}
s^\m= P\b^\m +\H^{\m\n}\b_\n-\a J^\m,
\end{eqnarray}
with $\b^\m=\b u^\m$ ($\b=1/T$), $\a=\m/T$, and $s^\m$ the entropy current such that $u\cdot s=s$. The divergence of $s^\m$ (multiplied by $T$) can be directly calculated,
\begin{eqnarray}
\label{entropy1}
T \pt_\m s^\m= \H^{\m\n}_{(1)}\nabla_\m u_\n - TJ^\m_{(1)}\nabla_\m\a.
\end{eqnarray}
The second law of local thermodynamics requires that $T\pt_\m s^\m\geq 0$ for any configurations of velocity field $u^\m$, temperature $T$, and chemical potential $\m$, which imposes the following constraints:
\begin{eqnarray}
\label{entropy2}
h_0=-\zeta\leq 0,\quad i_0=2\eta \geq 0,\quad J_{(1)}^\m=\s \nabla^\m\a,
\end{eqnarray}
where $\zeta$ and $\eta$ are the bulk and shear viscosities, and $\s$ the charge conductivity. This also shows that the coefficients $C'_0$ and $D'_0$ are fixed in such a way that $C'_0 \nabla^\m \ve + D'_0\nabla^\m n=\s\nabla^\m\a$. The EOMs of first-order dissipative hydrodynamics then read
\begin{eqnarray}
\label{navier1}
(\ve+P-\zeta\h)D u^\m-\nabla^\m (P-\zeta\h) +2\eta\Delta^\m_\n\pt_\r\s^{\n\r} = 0
\, ,
\\
\label{energyeom}
D\ve+(\ve+P-\zeta\h)\theta-2\eta\s_{\m\n}\s^{\m\n}
=0\,, \\
\label{chargeeom}
D n+n\h+\s\nabla^2\a=0.
\end{eqnarray}
The first equation is the relativistic Navier-Stokes equation. The above procedure can continue to higher order in gradients and will give us higher-order hydrodynamics. We, however, will not discuss these more complicated situation. The readers can find discussions in Refs.~\cite{Romatschke:2009im,Jeon:2015dfa,Yan:2017ivm,Florkowski:2017olj,Rocha:2023ilf}.

\section{Construction of relativistic spin hydrodynamics}\label{sec:spin}
With the above preparation, we now discuss the construction of relativistic spin hydrodynamics, in which the conservation of angular momentum is explicitly encoded within a (quasi)-hydrodynamic framework. The fundamental conservation laws are energy-momentum conservation (\ref{con:tmn}) and angular momentum conservation (\ref{con:Mmnr2}). Before delving into the detailed construction, we note that if we assign the spin density $S^{\m\n}=u_\r\S^{\r\m\n}$ as a dynamic variable in our framework, equation \eq{con:Mmnr2} means that it is generally not conserved. This reflects the fact that spin angular momentum can be transformed into orbital angular momentum, thus disqualifying it as a true hydrodynamic mode. Consequently, spin hydrodynamics will not be a strict hydrodynamic theory for gapless modes. Instead, it should be categorized as quasi-hydrodynamics, where the low-energy dynamic variables comprise true hydrodynamic modes and some gapped modes (quasi-hydrodynamic modes) whose gap in the low-momentum region is parametrically small compared to other microscopic modes (the hard modes of the system)~\cite{Hongo:2021ona}. This results in a spectrum separation; for physics at energy scales comparable to these modes, we can only consider the quasi-hydrodynamic modes alongside the true hydrodynamic modes. The so-called generalized hydrodynamics~\cite{Grozdanov:2018fic} and Hydro+~\cite{Stephanov:2017ghc} near the QCD critical point fall into this category. The spin hydrodynamics we are going to discuss also belongs to this type of theory. This framework requires that spin excitations, despite being gapped, remain low-energy excitations compared to other microscopic modes~\cite{Hongo:2021ona}. For instance, if the system contains massive fermions, the spins of these fermions are hard to relax, because the spin-orbit coupling is inversely suppressed by the mass of the fermions compared to typical energy transfer~\cite{Hongo:2022izs,Hidaka:2023oze,Li:2019qkf}. Thus, these spins are quasi-conserved and we can formulate a quasi-hydrodynamic theory for it, which is called the spin hydrodynamics.  

We consider a charge neutral system such as the quark gluon plasma or usual electric plasma in which some of constituent particles are spinful particles. The symmetry we are considering are the spacetime translation symmetry and Lorentz symmetry. They lead to the energy-momentum conservation and angular momentum conservation as given by \eq{con:tmn} and \eq{con:Mmnr2}. Now the spin tensor $\S^{\m\r\s}$ plays the role of the charge current $J^\mu$ and we could write it as $\S^{\m\r\s}=S^{\r\s} u^\m +{\rm higher\;\;order\;\; terms}$ with the spin density $S^{\r\s}$ playing similar role of charge density $n$ in \eq{j:exp5}. To proceed, we need to choose suitable power counting scheme for all the (quasi-)hydrodynamic variables. If we are considering the QGP in heavy ion collisions, from the measurements of global spin polarization of hyperons, we can know that the spin density in QGP should be small because the hyperon spin polarization is only about a few percent. Thus, it is reasonable to assume the spin density $S^{\r\s}$ to be parametrically smaller than the true hydrodynamic modes described by variables $\ve$ and $u^\m$. Thus, let us take the following power counting scheme:
\begin{eqnarray}
	\label{power:1}
&\ve, P, T, u^\m \sim O(1),&\\
&S^{\r\s}\sim O(\pt).&
\end{eqnarray}
Analogous to that chemical potential $\m$ is conjugate to charge density $n$, we can introduce the spin potential $\m^{\r\s}$ to be conjugate thermodynamically to spin density $S^{\r\s}$ and {\it propose} the first law for local thermodynamics as (analogous to \eq{ther1}):
\begin{eqnarray}
	\label{spinther1}
	& Tds +\frac{1}{2}\m_{\m\n} d S^{\m\n}= d\ve,&\\
	\label{spinther123}	&Ts+\frac{1}{2} \m_{\m\n} S^{\m\n} =\ve+P.&
\end{eqnarray}
Following the discussions about the fluid local frame, we realize that the same discussions are still valid for the symmetric part of the energy-momentum tensor and thus we still choose the definition of $u^\m$ such that it is the eigenvector of the symmetric part of the energy-momentum tensor, $\H^{\m\n}_{s}$ (we still call it the Landau-Lifshitz frame)~\footnote{Since $S^{\r\s}$ is counted as $O(\pt)$ quantities, the term $S^{\r\s}u^\m$ is unchanged at $O(\pt)$ under a re-definition of $u^\m\ra u^\m + \d u^\m$ with $\d u^\m\sim O(\pt)$. Therefore, \eq{llframespin} is automatically satisfied at $O(\pt)$ up on using the zeroth-order EOM for $u^\m$~\cite{Hattori:2019lfp}. But when there appear other conserved charges, such as a global $U(1)$ charge, \eq{llframespin} is a proposal to fix the local rest frame of the fluid.}:
\begin{eqnarray}
	\label{llframespin}
\H^{\m\n}_{s} u_\n=\ve u^\m.
\end{eqnarray}
Since the EOM for spin density involves only the antisymmetric part of the energy-momentum tensor, $\H^{\m\n}_{a}$, the symmetric part, $\H^{\m\n}_{s}$, still takes the same tensor decomposition upto the first order in gradients as \eq{tmn:exp5}:
\begin{eqnarray}
	\label{tmnsymm}
	\H^{\m\n}_{s} = \ve u^\m u^\n -(P-\z\h) \D^{\m\n} + 2\eta \s^{\m\n}+ O(\pt^2).	
\end{eqnarray}
To determine the form of $\H^{\m\n}_{a}$, we again ultilize the second law of local thermodynamics. The covariant entropy current reads (an analogue of \eq{gbrelation})
\begin{eqnarray}
	\label{gbrelationspin}
	s^\m= P\b^\m +\H^{\m\n}\b_\n-\frac{1}{2}\a_{\r\s} \S^{\m\r\s},
\end{eqnarray}
with $\a_{\r\s}=\m_{\r\s}/T$. The production rate of entropy then reads
\begin{eqnarray}
	\label{entropydivspin}
	T\pt_\m s^\m= \H^{\m\n}_{s(1)}\pt_{(\m} u_{\n)}+\H^{\m\n}_{a} \lb \m_{\m\n}+T\pt_{[\m} \b_{\n]}\rb +O(\pt^3).
\end{eqnarray}
The semi-positiveness of the first term on the right-hand side is already guaranteed when both bulk and shear viscosities are semi-positive. The requirement of the semi-positiveness of the second term gives the constitutive relation for $\H^{\m\n}_a$ at $O(\pt)$ order~\cite{Hattori:2019lfp}:
 \begin{eqnarray}
 	\label{tmnafirst}
 	\H^{\m\n}_{a} &=& q^\m u^\n -q^\n u^\m +\phi^{\m\n},\\
 	 	\label{tmnafirstq}
q^\m &=& \l\ls\b \nabla^\m T + D u^\m-2\m^{\m\n}u_\n\rs,\\
 	\label{tmnafirstphi}
 	\f^{\m\n} &=& \eta_s\D^{\m\r}\D^{\n\s} \lb \m_{\r\s}-T\varpi_{\r\s}\rb.
 \end{eqnarray}
The quantity 
 \begin{eqnarray}
	\label{thermvor}
\varpi_{\m\n}=(1/2)\lb\pt_\n\b_\m-\pt_\m\b_\n\rb
\end{eqnarray}
is called the thermal vorticity tensor. The quantities $\l$ and $\eta_s$ must be semi-positive to guarantee the semi-positiveness of entropy production. They are called boost heat conductivity and rotational viscosity, respectively~\cite{Hattori:2019lfp}. With these constitutive relations, we obtain the spin hydrodynamic equations up to $O(\pt^2)$ order:
\begin{eqnarray}
	\label{snavier1}
	(\ve+P-\zeta\h)D u^\m-\nabla^\m (P-\zeta\h) +2\eta\Delta^\m_\n\pt_\r\s^{\n\r} +q\cdot\pt u^\m-\D^\m_\n D q^\n-q^\m\h+\D^\m_\r\pt_\n\f^{\n\r}= 0
	\, ,\;\;\;\;\;
	\\
	\label{senergyeom}
	D\ve+(\ve+P-\zeta\h)\theta-2\eta\s_{\m\n}\s^{\m\n}+\pt\cdot q+q_\m D u^\m+\phi^{\m\n}\o_{\m\n}
	=0\,,\;\;\;\;\; \\
	\label{spineeom}
	D S^{\r\s}+S^{\r\s}\h+2\H^{\r\s}_a=0.\;\;\;\;\;
\end{eqnarray}

In this section, we present a detailed derivation of the constitutive relations up to first order in gradients for relativistic spin hydrodynamics. For related discussions following a similar approach, see, for example, Refs.\cite{Fukushima:2020ucl,She:2021lhe,Hongo:2021ona,Daher:2022xon,Cao:2022aku,Hu:2022azy,Biswas:2023qsw,Becattini:2023ouz,Drogosz:2024gzv,Dey:2024cwo,Yang:2024duc}. Other methodologies for deriving and analyzing the constitutive relations of spin hydrodynamics have also discussed in literature, including utilizing the hydrostatic partition function with constraints from the entropy current and Onsager relations~\cite{Gallegos:2021bzp,Gallegos:2022jow}, using local equilibrium and non-equilibrium statistical operators~\cite{Hu:2021lnx,Hu:2022azy,Tiwari:2024trl,Dey:2024cwo,She:2024rnx}, and employing kinetic theories~\cite{Florkowski:2017ruc,Florkowski:2018fap,Bhadury:2020puc,Shi:2020htn,Bhadury:2020cop,Peng:2021ago,Hu:2021pwh,Weickgenannt:2020aaf,Weickgenannt:2022zxs,Weickgenannt:2022qvh,Weickgenannt:2023btk,Drogosz:2024gzv,Wagner:2024fry}. Relativistic spin hydrodynamics has become a vibrant area of research, attracting intense discussions in recent years. In the following section, we will explore some of these developments; further insights can be found in, e.g., Refs.~\cite{Montenegro:2017rbu,Montenegro:2018bcf,Montenegro:2020paq,Li:2020eon,Wang:2021ngp,Wang:2021wqq,Torrieri:2022ogj,Hu:2022lpi,Hu:2022xjn,Sarwar:2022yzs,Daher:2022wzf,Shi:2023sxh,Wagner:2024fhf,Ren:2024pur,Wang:2024afv}.

\section{Discussions }\label{sec:disc}
We have developed spin hydrodynamics based on local thermodynamic laws. Spin hydrodynamics exhibits several novel features that differ significantly from those found in conventional relativistic hydrodynamics for other types of conservation laws (e.g., the energy-momentum conservation and baryon number conservation). In this section, we aim to explore and discuss some of these intriguing characteristics.

\subsection{Pseudo-gauge ambiguity}\label{sec:pseudo}
The definition of a conserved current is not unique. One example is the magnetization current and dipole charge density. Let $J^\mu = (\rho, \bm{J})$ represent the conserved conduction electric current. For a polarizable and magnetizable material, the total charge density and electric current are given by $\tilde{\rho} = \rho + \bm{\nabla} \cdot \bm{P}$ and $\tilde{\bm{J}} = \bm{J} + \bm{\nabla} \times \bm{M}$, respectively, where $\bm{P}$ is the electric dipole density, and $\bm{M}$ is the magnetization density. In covariant form, we have:
\begin{eqnarray}
\tilde{J}^\m = J^\m +\pt_\n {\cal M}^{\m\n}\quad {\rm with}\quad {\cal M}^{\m\n}=-{\cal M}^{\n\m}.
\end{eqnarray}
Obviously, the total current $\tilde{J}^\mu$ is conserved if the conduction current $J^\mu$ is conserved, and the total electric charge remains unchanged provided the surface dipole density vanishes. A transformation of a conserved current that preserves both the original conservation law and the total conserved charge is called a {\it pseudo-gauge transformation}. The above example demonstrates that the total current and the conduction current differ by a pseudo-gauge transformation (with the magnetization ${\cal M}^{\mu\nu}$ serving as the pseudo-gauge field). This example also highlights that a pseudo-gauge transformation is not a true gauge transformation, as it alters the physical content of the transformed current. Further insight about the pseudo-gauge transformation can be gained by examining the Maxwell equation:
\begin{eqnarray}
	\label{max1}
\pt_\m F^{\m\n}=\tilde{J}^\n.
\end{eqnarray}
One could subtract $-\pt_\r {\cal M}^{\n\r}$ from both sides and find
\begin{eqnarray}
\label{max12}
\pt_\m H^{\m\n} = J^\n,
\end{eqnarray}
where the new field strength tensor is defined as $H^{\m\n} \equiv F^{\m\n} + {\cal M}^{\m\n}$. This demonstrates that, without imposing additional constraints, the two sets of fields, $(F^{\m\n}, \tilde{J}^\m)$ and $(H^{\m\n}, J^\m)$, describe the same physical laws, and one can freely choose which set to use. (If further constraints are imposed---such as the Bianchi equation, $\pt_{[\m} F_{\m\n]} = 0$, which is not preserved under a general pseudo-gauge transformation---then only certain pseudo-gauges that respect the Bianchi equation are permitted.)

Similarly, let us consider angular momentum conservation (note the analogy with \eq{max12}, where $\S^{\m\n\r}$ and $\H^{\r\n} - \H^{\n\r}$ play roles analogous to $H^{\m\n}$ and $J^\m$ in \eq{max12}):
\begin{eqnarray}
	\label{con:Mmnr24}
	\pt_\m \S^{\m\n\r}=\H^{\r\n}-\H^{\n\r},
\end{eqnarray}
which is preserved under the transformation
\begin{eqnarray}
	\label{pseudog10}
\S^{\m\r\s} &\ra &\tilde{\S}^{\m\r\s}\equiv\S^{\m\r\s}-\Phi^{\m\r\s},\\
\H^{\m\n} &\ra&\tilde{\H}^{\m\n}\equiv \H^{\m\n}+\frac{1}{2}\pt_\l \Phi^{\l\m\n},
\end{eqnarray}
with $\Phi^{\l\m\n} = -\Phi^{\l\n\m}$ being an arbitrary local field. However, this transformation violates the conservation law of the energy-momentum tensor. It can be modified into the following form:
\begin{eqnarray}
	\label{pseudog1}
\S^{\m\r\s} &\ra &\tilde{\S}^{\m\r\s}\equiv\S^{\m\r\s}-\Phi^{\m\r\s},\\
\label{pseudog1tmn}
\H^{\m\n} &\ra&\tilde{\H}^{\m\n}\equiv \H^{\m\n}+\frac{1}{2}\pt_\l\lb \Phi^{\l\m\n}-\Phi^{\m\l\n}-\Phi^{\n\l\m}\rb,
\end{eqnarray}
which preserves both \eq{con:Mmnr24} and \eq{con:tmn}. Given a spacelike hypersurface $\Xi$, the total energy-momentum and total angular momentum across $\Xi$ are
\begin{eqnarray}
	\label{totaltmn}
P^\n &=& \int d\Xi_\m\H^{\m\n},\\
M^{\r\s} &=& \int d\Xi_\m M^{\m\r\s}= \int d\Xi_\m\lb x^\r\H^{\m\s}-x^\s\H^{\m\r}+\S^{\m\r\s}\rb. 
\end{eqnarray} 
One can check that $P^\m$ and $M^{\r\s}$ are invariant under pseudo-gauge transformation (\ref{pseudog1}) and (\ref{pseudog1tmn}) if the pseudo-gauge field $\Phi^{\m\r\s}$ vanishes at the boundary of $\Xi$~\footnote{This can be checked by noting that for $A^{\l\m\n}=-A^{\m\l\n}$ we have $\int d\Xi_\m\pt_\l A^{\l\m\n}=\int d\Xi_\m \pt^\perp_\l A^{\l\m\n}+\int d\Xi n_\m n_\l n\cdot\pt A^{\l\m\n}=\int d\Xi_\m \pt^\perp_\l A^{\l\m\n}$ with $n^\m$ the norm of $\Xi$ and $\pt^\perp_\l=\pt_\l-n_\l n\cdot\pt$. Then one can use the Gauss theorem to transform it to an integral over the boundary of $\Xi$.}.

One consequence of the existence of the pseudo-gauge transformation is the freedom to choose the symmetry properties of the spin tensor. To illustrate this, we consider an example in which we aim to transform a general spin tensor $\S^{\m\r\s} = -\S^{\m\s\r}$ into a completely antisymmetric form. We can choose $\F^{\m\r\s} = \S^{(\m\r)\s} - \frac{1}{2} \S^{\s\m\r}$. After applying the pseudo-gauge transformation, this yields:\begin{eqnarray}
	\label{pseudog3}
\S^{\m\r\s} &\ra &\tilde{\S}^{\m\r\s}
=\frac{1}{2}\lb \S^{\m\r\s}-\S^{\r\m\s}+\S^{\s\m\r}\rb,\\
	\label{pseudog333}
	\H^{\m\n} &\ra&\tilde{\H}^{\m\n}=\H^{\m\n}+\frac{1}{4}\pt_\l\lb 3\S^{\n\m\l}+\S^{\m\n\l}-\S^{\l\n\m}\rb.
\end{eqnarray}
Note that such-obtained $\tilde{\S}^{\m\r\s}$ is totally antisymmetric, so we can parameterize it as
\begin{eqnarray}
	\label{antispintensor}
	\tilde{\S}^{\m\r\s}=-\e^{\m\r\s\n} \tilde{S}_\n,
\end{eqnarray}
with $\tilde{S}^\m$ the corresponding spin (pseudo)vector. The spin density tensor is thus $\tilde{S}^{\m\n}=-\e^{\m\n\r\s}u_\r \tilde{S}_\s$. The main difference between this spin density tensor and the one used in Section \ref{sec:spin} is that $\tilde{S}^{\m\n}$ contains three degrees of freedom corresponding to the three spatial spin vectors, whereas $S^{\m\n}$ has six degrees of freedom, with three for spatial spin and three for boost. Thus, in some cases, it is more convenient to use $\tilde{\S}^{\m\r\s}$ to construct the spin hydrodynamics. By following a procedure similar to that adopted in Section \ref{sec:spin}, we can derive the constitutive relations in this context. In doing so, we decompose $\tilde{S}^\mu$ into $\tilde{S}^\m = \s^\m + n_5 u^\m$, where $\s^\m$ represents the spatial spin with the condition $\s \cdot u = 0$, and $n_5$ is a pseudoscalar field (hence the subscript $5$). We also decompose $\tilde{\H}^{\m\n}$ into:
\begin{eqnarray}
	\label{tmnantispin}
	\tilde{\H}^{\m\n} = \ve u^\m u^\n -P \D^{\m\n} + \tilde{\H}^{\m\n}_{s(1)}+\tilde{q}^\m u^\n -\tilde{q}^\n u^\m +\tilde{\phi}^{\m\n},
\end{eqnarray}
where, as we did in Section \ref{sec:spin}, we have assumed Landau-Lifshitz frame
\begin{eqnarray}
	\label{LLframe:new2}
\tilde{\H}^{\m\n}_s u_\n=\ve u^\m,
\end{eqnarray}
so that $\tilde{\H}^{\m\n}_{s(1)}$ is purely transverse to $u^\m$. It is important to note that, although we use the same symbols $\ve$, $P$, and $u^\m$ as in Section \ref{sec:spin}, their actual values may differ since the energy-momentum tensors and spin tensors in these two cases are different (but connected by the pseudo-gauge transformations (\ref{pseudog3}) and (\ref{pseudog333})). We adopt a power counting scheme similar to the one we chose in Section \ref{sec:spin}:
\begin{eqnarray}
	\label{power:new2}
	&\ve, P, T, u^\m \sim O(1),&\\
	&\tilde{S}^{\m}, \tilde{q}^\m, \tilde{\f}^{\m\n} \sim O(\pt).&
\end{eqnarray}
Using the same form for the entropy current and first law for local thermodynamics presented in Section \ref{sec:spin}, one can then find the divergence of the entropy current to be
\begin{eqnarray}
	\label{entropyprodnew}
	T\pt_\m s^\m= \tilde{\H}^{\m\n}_{s(1)}\pt_{(\m} u_{\n)}+\tilde{\H}^{\m\n}_{a} \lb \tilde{\m}_{\m\n}+T\pt_{[\m} \b_{\n]}\rb +O(\pt^3).
\end{eqnarray}
We note that in deriving this result, we have utilized the fact that contracting the equation of motion (\ref{con:Mmnr24}) with $u^\rho$ reveals that $\tilde{q}^\m$ is not an independent current, but is determined by $\tilde{S}^\m$ through the following relation:
\begin{eqnarray}
	\label{qmu}
\tilde{q}^\m=\frac{1}{2}\e^{\m\n\r\s}u_\n\nabla_\r\tilde{S}_\s.
\end{eqnarray}
This is because, when the spin tensor is completely antisymmetric, the components responsible for boost are gauged away, meaning that the corresponding torque for the boost in the antisymmetric part of the energy-momentum tensor cannot be an independent current either. Due to this relation, we can show that $n_5 = \tilde{S} \cdot u$ is actually an $O(\pt^3)$ quantity (and thus does not appear on the right-hand side of \eq{entropyprodnew}). In fact,  by direct calculation, one can find that the higher order terms that neglected in  \eq{entropyprodnew} contains only one term $\propto n_5$,
\begin{eqnarray}
\frac{1}{2}n_5 \e^{\m\n\r\s} u_\s\ls\pt_\m(\b\tilde{\m}_{\n\r})+\nabla_\m u_\r(\pt_\n\b+D\b_\n)\rs,
\end{eqnarray}
which infers that $n_5\propto \e^{\m\n\r\s} u_\s\ls\pt_\m(\b\tilde{\m}_{\n\r})+\nabla_\m u_\r(\pt_\n\b+D\b_\n)\rs \sim O(\pt^3)$ and thus can be neglected~\cite{Hongo:2021ona}. Therefore, from \eq{entropyprodnew}, we derive the constitutive relations for the spin hydrodynamics with a completely antisymmetric spin tensor as follows~\cite{Hongo:2021ona}:
\begin{eqnarray}
	\label{newconstrel}
\tilde{\H}^{\m\n}_{s(1)} &=& \z\h \D^{\m\n} + 2\eta \s^{\m\n}, \\
\tilde{\H}^{\m\n}_{a(1)} &=& \tilde{\f}_{(1)}^{\m\n}=\eta_s\D^{\m\r}\D^{\n\s} \lb \m_{\r\s}-T\varpi_{\r\s}\rb. 
\end{eqnarray}
Although these relations take the same form as those obtained in Section \ref{sec:spin}, it is important to note that they apply specifically to the pseudo-gauge of a totally antisymmetric spin tensor. These relations are particularly convenient for describing the evolution of spatial spin degrees of freedom.

Thus, we see that choosing different forms for the spin tensor (loosely referred to as different pseudo-gauges) leads to different forms for the constitutive relations. In an extreme case, one might even select $\F^{\m\r\s} = \S^{\m\r\s}$, which completely eliminates the spin tensor and renders the energy-momentum tensor totally symmetric (this choice is commonly referred to as the Belinfante gauge~\cite{Belinfante:1940,Rosenfeld:1940,Hehl:1976kj}). While this may seem to eliminate all information about spin in hydrodynamics, the energy density, viscous tensors, and heat current remain influenced by spin, meaning that the dynamics of spin are still embedded within those quantities. For discussions regarding the transformation from canonical to Belinfante gauges, see Refs.~\cite{Becattini:2018duy,Speranza:2020ilk,Fukushima:2020ucl,Li:2020eon,Daher:2022xon,Becattini:2023ouz}. Additionally, other pseudo-gauges have been employed and discussed in the context of spin hydrodynamics~\cite{Florkowski:2018fap,Bhadury:2020puc,Bhadury:2020cop,Bhadury:2021oat,Weickgenannt:2022jes,Singh:2024qvg,Buzzegoli:2024mra}.

\subsection{Spin hydrodynamics for strong vorticity}\label{sec:power}
The power counting scheme employed in the previous discussions is motivated by the observation that, at global equilibrium, the spin potential $\m_{\m\n}$ is determined by the thermal vorticity $\varpi_{\m\n}=(\pt_\n\b_\m-\pt_\m\b_\n)/2$, which is naturally assumed to be an $O(\pt)$ quantity. However, this assumption may not hold true because global equilibrium allows for arbitrarily large rotations (vorticity). When the vorticity is large, the assignment $\varpi_{\m\n}\sim O(\pt)$ becomes inadequate; instead, it is more appropriate to consider that $\varpi_{\m\n}\sim O(1)$. We will explore this situation in this subsection, following closely discussions in Ref.~\cite{Cao:2022aku}. Before going into the details, it is useful to compare spin hydrodynamics with magnetohydrodynamics (MHD), where the magnetic field is treated as an $O(1)$ quantity; See Ref.~\cite{Hattori:2022hyo} for a review of relativistic MHD.

MHD describes the coupled evolution of fluid energy-momentum (or temperature and velocity) and the electromagnetic field in the low-energy and long-wavelength regime. The fundamental equations consist of the conservation laws for the energy-momentum tensor and Maxwell's equations. Due to the screening effect, electric fields within the fluid are gapped and are parametrically small compared to the magnetic field. This renders the electric field not an active mode in the low-energy, long-wavelength regime. In contrast, there is no screening of the magnetic field, allowing it to exhibit its own dynamics even in this regime. Consequently, the magnetic field can be large and is treated as an $O(1)$ quantity, despite the fact that $\bm B=\bm\nabla\times\bm A$ involves one spatial gradient. The presence of an $O(1)$ magnetic field breaks the $SO(3)$ symmetry in the constitutive relations for $\H^{\m\n}$, introducing anisotropy even in ideal hydrodynamics. Specifically, we can define a normalized vector $b^\m=B^\m/B$, where $B=\sqrt{-B^\m B_\m}$, satisfying $b^2=-1$ and $b\cdot u=0$, as an additional building block for the hydrodynamic constitutive relations. For example, for a partiy-even and charge neutral fluid, the energy-momentum tensor can be decomposed into
\begin{eqnarray}
	\label{mhdtmn}
\H^{\m\n} &=& \ve u^\m u^\n -P_\perp\Xi^{\m\n}+P_\parallel b^\m b^\n +\H^{\m\n}_{(1)},
\end{eqnarray}
where $\Xi^{\m\n} = \D^{\m\n} + b^\n b^\n$ is a projector that is transverse to both $u^\m$ and $b^\m$. The terms $P_\perp$ and $P_\parallel$ represent the pressures in the directions transverse and parallel to the magnetic field, respectively. Note that when we allow an environmental parity violation (e.g., when there is a density imbalance between right- and left-hand particles in the fluid) and finite charge density, an term $u^{(\m}b^{\n)}$ can appear at zeroth order. The term $\H^{\m\n}_{(1)}$ (which is assumed to be symmetric since the spin degree of freedom is typically disregarded in MHD) denotes a collection of terms that are at least of order $O(\pt)$ in the gradient expansion and consistent with Onsager relations. For a parity-even fluid, all such terms are expressed as $\H^{\m\n}_{(1)} = \sum_{i=1}^7 \l_i \eta_i^{\m\n\r\s} \nabla_\r u_\s$, where $\l_i$ are the corresponding transport coefficients~\cite{Huang:2011dc,Grozdanov:2016tdf,Hernandez:2017mch}.
\begin{subequations}
\label{eq:kin:etatensor}
\begin{eqnarray}
\eta^{\m\n\rho\sigma}_1 &=& b^\m b^\n b^\rho b^\sigma,\\
\eta^{\m\n\rho\sigma}_2 &=& \Xi^{\m\n}\Xi^{\rho\sigma},\\
\eta^{\m\n\rho\sigma}_3 &=& -\Xi^{\m\n}b^\rho b^\sigma-\Xi^{\rho\sigma}b^\m b^\n,\\
\eta^{\m\n\rho\sigma}_4 &=& -2\left[b^{(\m}\Xi^{\n)\rho}b^\sigma +b^{(\m}\Xi^{\n)\sigma}b^\rho\right],\\
\eta^{\m\n\rho\sigma}_5 &=& 2\,\Xi^{\rho(\m}\Xi^{\n)\sigma}-\Xi^{\m\n}\Xi^{\rho\sigma},\\
\eta^{\m\n\rho\sigma}_6 &=& -b^{(\m}b^{\n)\rho}b^\sigma -b^{(\m}b^{\n)\sigma}b^\rho,\\
\eta^{\m\n\rho\sigma}_7 &=& \Xi^{\rho(\m}b^{\n)\sigma}+\Xi^{\sigma(\m}b^{\n)\rho},
\end{eqnarray}
\end{subequations}
where $b^{\m\n}=\e^{\m\n\r\s}u_\r b_\s$ is a cross projector which appears only when charge-conjugation symmetry is violated (e.g., when a net charge density is presented). 

Similar to the discussions above regarding MHD, we can consider a scenario for spin hydrodynamics where the vorticity is treated as zeroth order in gradients, while the gradients of other thermodynamic quantities are treated as first order. In line with MHD, this framework has been referred to as gyrohydrodynamics in Ref.~\cite{Cao:2022aku}. To simplify the notation, we reuse $b^\m$ to denote the unit vector along the vorticity:
\begin{eqnarray}
b^\m=\varpi^\m/\sqrt{-\varpi_\m\varpi^\m}=\o^\m/\sqrt{-\o_\m\o^\m},
\end{eqnarray}
with $\varpi^\m=-\e^{\m\n\r\s}u_\n\pt_\r\b_\s/2=\b\o^\m$ the thermal vorticity vector. We choose the pesudo-gauge so that the spin tensor is totally antisymmetric. Using $u^\m, b^\m$ as well as $g^{\m\n}, \e^{\m\n\r\s}$ as building blocks, we can decompose $\H^{\m\n}$ and $\S^{\m\n\r}$ into the following irreducible forms:
\begin{eqnarray}
\label{eqngyrohydro}
  \Theta^{\mu\nu}&=& \ve u^\mu u^\nu - P_\perp \Xi^{\m\n}+P_\parallel b^\m b^\n+P_\times b^{\m\n}+ q^\mu u^\nu - u^\mu q^\nu+\H^{\m\n}_{s(1)}+\f^{\m\n},\\
\Sigma^{\mu\nu\rho}&=&- \epsilon^{\mu\nu\rho\lambda} S_\l = -\epsilon^{\mu\nu\rho\lambda} 
(n_5 u_\lambda - S_\parallel b_\l+S_{\perp\lambda}),
\end{eqnarray}
where $P_{\perp,\parallel,\times}$ represent pressures (which will be counted as $O(1)$ quantities in gradient expansion) in different directions, whose physical meaning will become clear shortly. The quantity $S_\parallel=b\cdot S$ denotes the spin component in the direction of the vorticity, while $S_{\perp}^\m=\Xi^{\m\n} S_\n$ denotes the spin component transverse to the vorticity. As before, we have chosen the Landau-Lifshitz frame, with $q^\m$, $\H^{\m\n}_{s(1)}$, $\f^{\m\n}$, and $S_{\perp}^\m$ all transverse to $u^\m$. Note again that, with this fully antisymmetric choice of spin tensor, the $q^\m$ vector is no longer independent but is determined by $S^\m$ through \eq{qmu}.

The power counting scheme is such that $S_\parallel$ is counted as order one, while $\f^{\m\n} = -\f^{\n\m}$, $S_{\perp}^\m$, $n_5$, and $q^\m$ (see \eq{qmu}) are counted as being at least $O(\pt)$. Additionally, we will count $S^\m$ as $O(\hbar)$ (since spin is totally quantum in nature) in comparison to other thermodynamic quantities, which can appear even at the classical level and are therefore assigned $O(\hbar^0)$. This allows for a double expansion in both $\pt$ and $\hbar$. For the entropy current, we can write $s^\m =s u^\m+s^\m_{(1)}$ and use still \eq{spinther1}. It is straightforward to derive the divergence of the entropy current, and after some calculations, one finds that up to $O(\hbar\pt^2,\pt^3)$~\cite{Cao:2022aku}:
\begin{eqnarray}
 \partial_\mu s^\mu
&=& 
\left[ s - \beta \left( \ve + P_\perp \right) \right] 
\theta - ( P_\parallel -P_\perp- \mu_{\parallel} S_{\parallel} ) b^\mu b^\nu
\partial_\mu \beta_\nu
+ P_\times b^{\mu\nu} ( \partial_\mu \beta_\nu + \beta \mu_{\mu\nu} )
\nonumber \\
&&\quad 
+\Theta^{\mu\nu}_{s(1)} \partial_{(\mu} \beta_{\nu)}
+\f^{\mu\nu} ( \partial_{[\mu} \beta_{\nu]} + \beta \mu_{\mu\nu} )
+ \partial_\mu 
\left( 
 s_{(1)}^\mu 
- \beta \mu^\mu n_5 
\right)
+ O (\hbar \partial^2, \partial^3) .
\end{eqnarray}
The first line gives the zero-order contribution to the entropy production which is expected to vanish so that they represent the non-dissipative contributions. This gives 
\begin{eqnarray}
\ve+P_\perp = Ts\,,\;\;\; P_\parallel = P_\perp+\m_\parallel S_\parallel\,,\;\;\; P_\times=0.
\end{eqnarray}
The first relation is simply the Gibbs-Duhem relation, indicating that $P_\perp$ can be interpreted as the thermodynamic pressure. The second relation shows that the pressure along the vorticity direction differs from the thermodynamic pressure by an amount due to spin polarization $\m_\parallel S_\parallel$. This term is very similar to the $MB$ term in magnetohydrodynamic constitutive relation. The third relation shows that at leading order there is no spin torque. 

At $O(\pt)$ order, the requirement of a semi-positive entropy production gives that
\begin{eqnarray}
\label{strongvc1}
 	\Theta^{\mu\nu}_{s(1)}
	&=&  T  \eta^{\mu\nu\rho\sigma}   \partial_{(\rho} \beta_{\sigma)}
	+ T   \xi^{\mu\nu\rho\sigma}  
	( \partial_{[\rho} \beta_{\sigma]} + \beta \mu_{\rho\sigma} ) 
	,
	\\
\label{strongvc2}	\f^{\mu\nu}
	&=&  T   \gamma^{\mu\nu\rho\sigma}    ( \partial_{[\rho} \beta_{\sigma]} + \beta \mu_{\rho\sigma} ) 
	+ T \xi^{\prime \, \mu\nu\rho\sigma}  \partial_{(\rho} \beta_{\sigma)}    ,
	\\
	s_{(1)}^\mu
	&=& 
	 \beta \mu^\mu n_5,
\end{eqnarray}
where $\eta^{\m\n\r\s}$ and $\g^{\m\n\r\s}$ are the usual and rotational viscous tensors representing the response of the symmetric and antisymmetric parts of the energy-momentum tenor to fluid shear and expansion and the difference between vorticity and spin potential, and $\xi^{\m\n\r\s}$ and $\xi^{\prime\,\m\n\r\s}$ are two cross viscous tensors. Note that the cross viscous tensors are not independent from each other but inter-related according to Onsager's reciprocal principle, $\xi^{\prime \, \mu\nu\rho\sigma} (b) =  \xi^{\rho\sigma\mu\nu} (-b)  $. Decomposing these tensors into irreducible structures, one obtains a number of new transport coefficients (viscosities) that characterize the response of the fluid to gradients of fluid velocity and spin potential~\cite{Cao:2022aku}:
\begin{eqnarray}
  \eta^{\mu\nu\rho\sigma}
&=& \zeta_{\perp} \Xi^{\mu\nu} \Xi^{\rho\sigma}
+ \zeta_{\parallel} b^\mu b^\nu b^{\rho} b^\sigma
+ \zeta_{\times} 
\big( b^\mu b^\nu \Xi^{\rho\sigma} + \Xi^{\mu\nu} b^\rho b^\sigma
\big)
\non
&&\quad 
+ \eta_{\perp} 
\big( 
\Xi^{\mu\rho} \Xi^{\nu\sigma} 
+ \Xi^{\mu\sigma} \Xi^{\nu\rho} 
- \Xi^{\mu\nu} \Xi^{\rho\sigma}  
\big)
+ 2 \eta_{\parallel} 
\big( b^\mu \Xi^{\nu(\rho} b^{\sigma)}
+ b^\nu \Xi^{\mu(\rho} b^{\sigma)} 
\big)
\non
&&\quad 
+ 2 \eta_{H_\perp} 
\big( 
\Xi^{\mu(\rho} b^{\sigma)\nu}
+ \Xi^{\nu(\rho} b^{\sigma)\mu}
\big)
+ 2 \eta_{H_\parallel} 
\big( 
b^\mu b^{\nu(\rho} b^{\sigma)}
+ b^\nu b^{\mu(\rho} b^{\sigma)}
\big),
\\
\gamma^{\mu\nu\rho\sigma}
&=& \gamma_\perp 
\big(
\Xi^{\mu\rho} \Xi^{\nu\sigma} - \Xi^{\mu\sigma} \Xi^{\nu\rho}
\big)
+ 2 \gamma_{\parallel}   
\big( 
b^\mu \Xi^{\nu[\rho} b^{\sigma]}
-  b^\nu \Xi^{\mu[\rho} b^{\sigma]} 
\big)
\non
&&\quad 
+ 2  \gamma_{H } 
\big( 
b^\mu b^{\nu[\rho} b^{\sigma]}
- b^\nu b^{\mu[\rho} b^{\sigma]}
\big),
\\
\xi^{\mu\nu\rho\sigma} 
&=&  
2 \xi_\parallel \big( b^\mu \Xi^{\nu [\rho} b^{\sigma]} + b^\nu \Xi^{\mu [\rho} b^{\sigma]}
\big)
+  \zeta_{H_\perp}   
  \Xi^{\mu\nu}  b^{\rho\sigma} 
+  \zeta_{H\parallel}  b^\mu b^{\nu}  b^{\rho\sigma}
\non
&&\quad
+ 2 \xi_{H}   \big(   b^\mu b^{\nu[\rho} b^{\sigma]}  + b^\nu b^{\mu[\rho} b^{\sigma]} 
\big)
,
\end{eqnarray}
where the $\eta$'s, $\z$'s, $\g$'s, and $\xi$'s are transport coefficients. Especially, those with subscript ``H" are Hall-type transport coefficients which do not contribute to the entropy production and thus their sign are not constrained by the second law of local thermodynamics. One may wonder why  term $\propto b^{\m\n} b^{\r\s}$ (such term would contribute to an $O(\partial)$ analogue of $P_\times$ term in \eq{eqngyrohydro} ) does not appear in $\g^{\m\n\r\s}$. This is because it is actually not independent from other terms in $\g^{\m\n\r\s}$~\cite{Huang:2011dc}. Note also that the expression for $\xi^{\m\n\r\s}$ is different from that in Ref.~\cite{Cao:2022aku} but equivalently gives the same constitutive relations once substituting into \eq{strongvc1}.

\subsection{A spin Cooper-Frye formula}\label{sec:cf}
For the purpose of applying spin hydrodynamics to specific physical systems, we need to know what are the appropriate observables for detection of the spin degrees of freedom in the fluid. In principle, the presence of the spin degree of freedom in the fluid should modify the usual hydrodynamic quantities such as the energy density and fluid velocity, but when the spin density is not large (nevertheless it is always suppressed by $\hbar$ comparing to the traditional hydrodynamic quantities), such modification is small. In heavy ion collisions, the natural observable is the spin polarization of hadrons, including spin-1/2 hyperons and spin-1 vector mesons. Hyperons are of special interest because they primarily decay via weak interactions in such a way that the momentum of one of the daughter particles tends to align with the spin direction of the hyperon. In order to obtain the spin polarization observables of hadron from the spin hydrodynamics, we need a machinery to convert the outcomes of spin hydrodynamics, such as fluid velocity, temperature, and spin potential, to measurable hadronic observables.

In the application of traditional hydrodynamics to heavy-ion collisions, the hadron momentum spectra are typically obtained using the so-called Cooper-Frye formula,
\begin{eqnarray}
	\label{cooperfusual}
E_p\frac{d N_i}{d\bp^3} = \int_\Xi d\Xi_\m(x) p^\m f_i(x,p),
\end{eqnarray}
where the integral is over the freeze-out hypersurface (where particlization occurs) $\Xi$,  and $f_i(x,p)$ is the distribution function of species $i$ of the hadrons in the fluid. Any possible degeneracy of hadrons should be accounted for in $f_i$. For example, when dissipative effects are neglected, the distribution function $f_i$ is typically taken as the Fermi-Dirac or Bose-Einstein functions $f_{F,B}(p\cdot \b-\m_i)$ with $\m_i$ the chemical potential. The above Cooper-Frye formula has been widely used in hydrodynamic simulations in heavy-ion collisions and has proven to be very successful. Therefore, in extending traditional hydrodynamics to spin hydrodynamics, we also need to generalize the above Cooper-Frye formula to a spin Cooper-Frye formula. 

Let us consider a system in which the thermal equilibrium is reached locally but not necessarily globally. The density operator $\hat\r$ for description of such an ensemble is obtained by maximizing the entropy functional under the constraints of given energy-momentum and angular momentum (or spin) densities,
\begin{eqnarray}
	\label{entropyf}
S[\hat\r]&=&-\Tr (\hat\r\ln\hat\r) + \l(\Tr\hat\r -1)-\int_\Xi d\Xi_\m \big[\Tr(\hat\r\hat{\H}^{\m\n})-\H^{\m\n}\big]\b_\n\non&& + \frac{1}{2}\int_\Xi d\Xi_\m \big[\Tr(\hat\r\hat{\S}^{\m\n\r})-\S^{\m\n\r}\big]\m_{\n\r} ,
\end{eqnarray}
where $\H^{\m\n}(x)$ and $\S^{\m\n\r}(x)$ are the actual local energy-momentum tensor and spin tensor, and $\b_\n(x)$ and $\m_{\n\r}(x)$ are corresponding Lagrange multipliers. The Lagrange multiplier $\l$ is introduced to normalize $\hat\r$ and is related to the partition function $Z$ as $\exp(1-\l)=Z$. The resultant density operator is the local-equilibrium density operator~\cite{Zubarev:1979afm,vanWeert:1982,Becattini:2014yxa,Becattini:2019dxo}:
\begin{eqnarray}
	\label{ledo}
\hat\r_{\rm LE} =\frac{1}{Z_{\rm LE}}\exp\lc-\int_\Xi d\Xi_\m(x) \ls \hat\H^{\m\n}(x)\b_\n(x)-\frac{1}{2}\hat{\S}^{\m\r\s}(x)\m_{\r\s}(x)\rs\rc,
\end{eqnarray}
where $Z_{\rm LE}$ is the local-equilibrium partition function. Now, we see that $\hat{\rho}_{\rm LE}$ is determined by the local thermodynamic quantities $\beta^\mu$ and $\mu_{\rho\sigma}$. If we calculate the spin density $\Sigma^{\mu\rho\sigma}(x)$ using this density operator, we obtain a relation between $\Sigma^{\mu\rho\sigma}(x)$ and the local thermodynamic quantities (and possibly their derivatives). However, this is not particularly useful in the context of heavy-ion collisions, because what is measured is the spin density in momentum space, rather than coordinate space. To express such a relation for the spin density in momentum space or phase space, the most natural approach is to use the Wigner function.

To illustrate how this can be achieved, we consider a system of Dirac fermions as an example. The Wigner operator is defined as
\begin{eqnarray}
\label{Wigneroperator}
\hat{W}(x,p)&=&\int d^4s  e^{-ip\cdot s} \bar{\hat\j}\left(x+\frac{s}{2}\right) \otimes \hat{\j} \left(x-\frac{s}{2}\right),\;\;\;\;
\end{eqnarray}
where $[\bar{\hat\psi}\otimes\hat{\psi}]_{ab}\equiv\bar{\hat\psi}_b\hat{\psi}_{a}$ with $a,b$ spinor indices. We choose the canonical pseudo-gauge in which the energy-momentum tensor operator and spin tensor operator are given by 
\begin{eqnarray}
	\label{fcanonicaltmn}
\hat{\H}^{\m\n} &=&\bar{\hat\j}i\g^\m\pt^\n\hat\j -\eta^{\m\n}\hat{\cal L} ,\;\\
\hat\S^{\m\n\r} &=& \frac{1}{4}{\bar{\hat\j}} \lc \g^\m, \s^{\r\s} \rc \hat{\j} = -\frac{1}{2} \e^{\m\n\r\s}\bar{\hat\j}\g_\s\g_5\hat\j.
\end{eqnarray}
where $\hat{\cal L}$ is the Lagrangian (in the following we consider free fermions, so that $\cl=\bar\j\lb i\g^\m\pt_\m - m\rb\j$ is in quadratic form and the second term in $\hat{\H}^{\m\n}$ vanishes upon using equation of motion of field operator) and $\s^{\r\s}=i[\g^\r, \g^\s]/2$. Note that the second equation means that the spin vector $\hat S_\s = (1/2)\bar{\hat\j}\g_\s\g_5\hat\j$ is nothing but half of the axial current. In the above, both $\hat{\H}^{\m\n}(x) $ and $\hat\S^{\m\n\r}(x)$ are local Heisenberg operators. We can extend them into operators in phase space by using the Wigner transformation, e.g.,
\begin{eqnarray}
	\label{fcanonicalphasespace}
\hat\S^{\m\n\r} (x,p)&=&-\frac{1}{2}\e^{\m\n\r\s}\int d^4s  e^{-ip\cdot s} \bar{\hat\j}\left(x+\frac{s}{2}\right)\g_\s\g_5\hat{\j} \left(x-\frac{s}{2}\right)\non
&=&-\frac{1}{2}\e^{\m\n\r\s} \Tr_{\rm D} \ls \g_\s\g_5 \hat{W}(x,p) \rs\,,
\end{eqnarray}
where $\Tr_{\rm D}$ is trace over Dirac space. It is easy to find that $\int d^4 p/(2\p)^4\hat\S^{\m\n\r} (x,p)=\hat\S^{\m\n\r} (x)$. The integration of $\hat\S^{\m\n\r} (x,p)$ over certain spacelike hypersurface will give us the spin tensor in momentum space (whose exact meaning will be clarified later) whose ensemble average under ${\hat\r}_{\rm LE}$ is exactly the quantity that we are looking for. Therefore, we just need to calculate the so-called Wigner function under local equilibrium:
\begin{eqnarray}
\label{wignerfunc}
W(x,p)= \lan \hat{W}(x,p) \ran = {\rm Tr} \ls \hat{\r}_{\rm LE}\hat{W}(x,p) \rs,
\end{eqnarray}
where ${\rm Tr}$ denotes the trace over a complete set of microstates of the system. To proceed, one can re-write the local-equilibrium density operator as $\hat{\r}_{\rm LE}=\exp(\hat A +\hat B)/Z_{\rm LE}$ with the abbreviations
\begin{eqnarray}
	\hat{A}&=&-\hat{P}^\m\b_\m(x),\\
	\hat{B}&=&-\int d\Xi_\m(y) \big[ \hat{\H}^{\m\n}(y)\D\b_\n(y)-\frac{1}{2}\hat{\S}^{\m\r\s}(y)\m_{\r\s}(y)\big],
\end{eqnarray}
where $\hat{P}^\m=\int d\Xi_\n(y)\hat{\H}^{\n\m}(y)$, $\D\b_\m(y)=\b_\m(y)-\b_\m(x)$. 
The purpose of rewriting $\hat{\rho}_{\rm LE}$ in this form is that, typically, the correlation length between the spin tensor and the energy-momentum tensor is small. Within this correlation length, we can assume that local thermodynamic quantities, such as $\beta_\mu$, vary only slightly. Given that $\mu_{\rho\sigma}$ is also small at the hypersurface $\Xi$ (which is a reasonable assumption for heavy-ion collisions, though it may not hold for a strongly polarized medium), we assign $\Delta \beta_\nu \sim \mu_{\rho\sigma} \sim O(\pt)$, and therefore $\hat{A} \sim O(1)$, $\hat{B} \sim O(\pt)$. Using this power-counting scheme, we can expand the right-hand side of \eq{wignerfunc} order by order in $\pt$ by applying the identity $
e^{\hat{A}+\hat{B}}= e^{\hat{A}}+ e^{\hat{A}} \int^1_0 d\l e^{-\l \hat{A}} \hat{B}e^{\l \hat{A}} + \cdot\cdot\cdot $,  and obtain
\begin{eqnarray}
	\label{ensembleave3}
	W(x,p) &=& W_0(x,p) + W_1(x,p)+\cdots\,,
\end{eqnarray}
where
\begin{eqnarray}
	\label{ensembleave4}
	W_0(x,p) &=& \langle \hat{W}(x,p) \rangle_{0} \equiv \frac{1}{Z_0}\Tr\lb e^{\hat{A}}\hat{W}(x,p)\rb,\qquad\\
	W_1(x,p) &\equiv& \lan\hat{W}(x,p)\ran_{(\H)} + \lan\hat{W}(x,p)\ran_{(\S)}\,\,,
\end{eqnarray}
with
\begin{equation}
	\begin{split}
		\lan\hat{W}(x,p)\ran_{(\H)} &\equiv - \int_0^1d\l \int d\Xi_{\r}(y)\D\b_{\n}(y)\lan\hat{\H}^{\r\n}(y-i\l \b(x))\hat{W}(x,p)\ran_{0,c}\,,\quad\\
		\lan\hat{W}(x,p)\ran_{(\S)} &\equiv  \frac{1}{2}\int_0^1d\l \int d\Xi_{\n}(y)\m_{\r\s}(y)\lan\hat{\S}^{\n\r\s}(y-i\l \b(x))\hat{W}(x,p)\ran_{0,c}\,,
		\label{OTJS}
	\end{split}
\end{equation}
and $Z_0=\Tr e^{\hat{A}}$. Here, $\lan\cdots\ran_{0,c}$ means the connected part of the correlation. The calculation then will depend on the shape of the hypersurface $\Xi$. For the purpose of illustration, we consider $\Xi$ to be the 3-space at some time $t$ so that its normal direction is $\hat{t}^\m=(1,\bm 0)$. The calculation then is straightforward by using the free field operator
\begin{equation}
\begin{split}
\hat{\j}(x)&=\sum^2_{\s=1}\frac{1}{(2\p)^{3/2}} \int \frac{d^3\bk}{2E_k} \big[ u_\s(\bk) e^{-ik\cdot x} \hat{a}_\s(\bk) +  v_\s(\bk) e^{ik\cdot x} \hat{b}^\dag_\s(\bk) \big]\,,
\end{split}
\end{equation}
where $E_k=\sqrt{\bk^2+m^2}$ and $\hat{a}_\s(\bk),\hat{b}_\s(\bk)$ are annihilation operators for particles and antiparticles satisfing the anti-commutation relation $\{ \hat{a}_\s(\bk), \hat{a}^\dag_{\s'}(\bq) \}=\{ \hat{b}_\s(\bk), \hat{b}^\dag_{\s'}(\bq) \} = 2E_k \d_{\s\s'}\d^3(\bk-\bm{q})$ and the relation $\lan \hat{a}^\dag_\s(\bk) \hat{a}_{\s'}(\bq)\ran_0=\lan \hat{b}^\dag_\s(\bk) \hat{b}_{\s'}(\bq)\ran_0=2E_k\d_{\s\s'}\d^3(\bk-\bq) n_F(k\cdot\b)$. In the following, we consider only the particle branch, the antiparticle branch is completely similar. The zeroth order Wigner function is easy to get: $W_0(x,p)=2\p (p\!\!\!/+m)\h(p_0)\d(p^2-m^2)n_F(p\cdot\b)$, which is spin independent: $\Tr_D \ls \g^\m\g_5 W_0(x,p)\rs = 0$. 

The first-order Wigner function reads 
\begin{eqnarray}
\label{Wtsj}
\lan \hat{W}(x,p)\ran_{(\H/\S)}
&=& 2\p\int_0^1 d\l \int \frac{d^3\bk}{2E_k} \int \frac{d^3\bq}{2E_q}
 \d^4\lb p-\frac{q+k}{2} \rb (\g\cdot k+m)\hat{t}_\m I_{(\H/\S)}^\m (\g\cdot q+m) \non
&&\times e^{\l(k-q)\cdot\b(x)} n_F(k)\ls 1 - n_F(q) \rs\,,
\end{eqnarray}
where $n_F(p)=n_F[\b(x)\cdot p]$, $I_{(\H)}^\m = - \g^\m p^\n\ls\pt_\l\b_\n(x)\rs\D^\l_\b \ls i\pt_q^\b \d^3(\bm{q}-\bm{k}) \rs$, and $I_{(\S)}^\m = \frac{1}{4}\e^{\m\n\r\s} \g^5\g_\n \m_{\r\s} \d^3(\bm{q}-\bm{k})$ with $\D^{\m\n}=\w^{\m\n}-\hat{t}^\m\hat{t}^\n$. To obtain this result, we have used 
\begin{equation}
\begin{split}
\int d\Xi_{\m}(y)(y-x)^{\a}e^{-i(p-q)\cdot(y-x)} = (2\p)^3
\hat{t}_\m \D^{\a}_{\b}\frac{i\pt}{\pt p_{\b}}\d^{3}(\bp-\bq),
\end{split}
\end{equation}
which is valid when $\Xi$ is a 3-space. In heavy-ion collisions, the true freeze-out hypersurface $\Xi$ is of course not a 3-space and thus correction due to the non-flatness of $\Xi$ would appear; see discussions in Refs.~\cite{Becattini:2021suc,Sheng:2024pbw}.

With the first-order Wigner function in Eq.~\eqref{Wtsj}, the local-equilibrium spin vector in phase space is directly obtained by finishing the trace over Dirac space~\cite{Liu:2021nyg,Buzzegoli:2021wlg}:
\begin{eqnarray}
\label{doa}
S_\m(x,p) &=- 4\p \d( p^2-m^2 )\h(p_0) n_F(p) [ 1 - n_F(p) ] \Big\{ \frac{1}{4} \e_{\m\n\r\s} p^\n \m^{\r\s}+  \S^{\hat{t}}_{\m\n} \ls (\x^{\n\l}+\D\m^{\n\l}) p_\l \rs \Big\},\;\;\;\;\;\;\;
\end{eqnarray}
where $\S^{\hat t}_{\m\n}=\e_{\m\n\r\s}p^\r \hat{t}^\s/(2p\cdot\hat{t})$, $\xi_{\m\n}=\pt_{(\m}\b_{\n)}$ is the thermal shear tensor, and $\D\m^{\m\n}=\m^{\m\n}-\varpi^{\m\n}$ is the difference between spin potential and thermal vorticity tensor. 

With this spin vector in phase space, the spin vector per particle in momentum space is obtained by average over hypersurface $\Xi$~\cite{Liu:2021nyg,Buzzegoli:2021wlg}:
\begin{eqnarray}
\label{spincooperfrye}
\displaystyle S_\m(p) &=&\frac{1}{2}\frac{\int d\Xi(x)\cdot p\Tr_D[\g^\m\g^5W(x,p)]}{\int d\Xi\cdot p\Tr_D[ W(x,p)]}\non
\displaystyle&=& - \frac{\int d\Xi\cdot p \lc \e_{\m\n\a\b} p^\n \m^{\a\b}
+ 4\S^{\hat t}_{\m\n} \ls p_\l (\x^{\n\l}+\D\m^{\n\l})\rs \rc n_F ( 1 - n_F )}{8m\;\int d\Xi\cdot p\; n_F},
\end{eqnarray}
where $p^\m$ on the right-hand side is on-shell. This is a Cooper-Frye-type formula for the spin vector, which connects the momentum-space distribution of the mean spin vector of particles emitted from $\Xi$ with the fluid properties characterized by $\mu^{\mu\nu}(x)$ and $\beta^\mu(x)$ on $\Xi$. Thus, once these fluid variables are obtained from spin hydrodynamics, this spin Cooper-Frye formula allows us to convert them into the mean spin vector in momentum space, which is a directly measurable quantity. 

We give several comments before concluding this subsection. First, at local equilibrium, the thermal shear tensor can induce spin polarization, which has important implications for spin polarization phenomenology in heavy-ion collisions~\cite{Becattini:2021suc,Liu:2021uhn,Becattini:2021iol,Fu:2021pok}. Second, when the system is in global equilibrium, the spin potential is determined by the thermal vorticity, and the thermal shear tensor $\xi_{\mu\nu}$ vanishes. In this case, the above spin Cooper-Frye formula reduces to the one obtained in Refs.~\cite{Becattini:2013fla,Fang:2016vpj,Liu:2020flb}. Third, we have not included the effect of a finite baryon chemical potential. Its inclusion is straightforward, with the modification being that the distribution function $n_F(p \cdot \beta) \to n_F(p \cdot \beta - \alpha)$, where $\alpha = \mu/T$. Additionally, a new term $4 \int d\Xi , p \cdot \Sigma^{\hat{t}}_{\mu\nu} \partial^\nu \alpha$ should be added to the numerator of \eq{spincooperfrye}, whose contribution is referred to as the spin Hall effect~\cite{Liu:2020dxg}. Fourth, the formula (\ref{spincooperfrye}) depends on the choice of the pseudo-gauge~\cite{Liu:2021nyg,Buzzegoli:2021wlg}. In particular, it is possible to completely eliminate the contributions from thermal shear by adopting appropriate pseudo-gauges. Therefore, when applying this formula to spin hydrodynamics, it is important to be careful in choosing the pseudo-gauge to maintain consistency.

\section{Summary and outlooks}\label{sec:summ}
In this article, we provide a pedagogical introduction to relativistic spin hydrodynamics. We begin by demonstrating how one can derive a set of hydrodynamic equations from the conservation equations based on the requirement of local thermodynamic laws, primarily the second law of thermodynamics. We then extend this framework to include the conservation of angular momentum, which leads to spin hydrodynamics. In the framework of spin hydrodynamics, the new (quasi-)hydrodynamic variable is the spin density. Due to spin-orbit coupling, the spin density is not a strict hydrodynamic variable but rather a quasi-hydrodynamic variable. It relaxes to a local equilibrium value determined by the local thermal vorticity, through the dissipative conversion of spin and orbital angular momenta. We show how such dissipative processes are characterized by two new transport coefficients: one for boost heat conductivity and another for rotational viscosity.

We discuss several interesting aspects of spin hydrodynamics. First, we address the pseudo-gauge ambiguity in defining the spin tensor, which reflects the freedom in separating the total angular momentum into its spin and orbital components. One consequence of this pseudo-gauge ambiguity is that we have the flexibility to choose spin tensors with different symmetries in their indices as the starting point for derivation of the spin hydrodynamics, leading to different constitutive relations. Second, we emphasize the importance of derivative power counting in the formulation of spin hydrodynamics. In particular, for a strongly vortical (or strongly spin-polarized) fluid, it is natural to assign the vorticity and the spin potential as being of similar strength to other local thermodynamic quantities, such as temperature, in terms of derivative powers. This is analogous to magnetohydrodynamics. As a result, anisotropy emerges in the constitutive relations both at zeroth order and first order in derivatives. This framework is well-suited for describing strongly vortical or spin-polarized fluids. Third, for potential applications of spin hydrodynamics, such as in heavy-ion collisions, we require a method to convert the results of spin hydrodynamics---specifically, the spin density (or spin potential), temperature, and fluid velocity---into momentum-space observables. To this end, we give a spin Cooper-Frye formula for Dirac fermions, and a similar formula can also be derived for spin-one vector bosons.

Spin hydrodynamics is an area of intensive study, with many interesting aspects already explored and many more awaiting investigation. We provide a brief discussion of some of these topics.

{\bf (1) Spin magnetohydrodynamics.} When the constituents of the fluid are charged, the fluid can interact with electromagnetic fields and behave like a magnetized fluid. In this case, it is convenient to extend spin hydrodynamics to spin magnetohydrodynamics~\cite{Singh:2022ltu,Bhadury:2022ulr,Kiamari:2023fbe,Fang:2024sym,Fang:2024hxa}. Since electric fields are easily screened, they are not typically described as hydrodynamic variables. Therefore, the new hydrodynamic variable is the magnetic field (more precisely, the magnetic flux) $B^\m = \tilde{F}^{\m\n}u_\n$, which is counted as an $O(1)$ quantity in derivative power counting. Its conservation law is simply the Bianchi identity.
\begin{eqnarray}
	\label{mhdspin1}
\pt_\m\tilde{F}^{\m\n}=0.
\end{eqnarray}
Here, $\tilde{F}^{\m\n}=(1/2)\e^{\m\n\r\s}F_{\r\s}$ is the dual tensor of Maxwell tensor. This equation should be combined with the conservation laws of energy-momentum and angular momentum to form the complete equations of motion for the fluid. Expanding $\tilde{F}^{\m\n}$ in terms of the hydrodynamic variables gives~\cite{Hattori:2022hyo}
\begin{eqnarray}
	\label{mhdexp}
\tilde{F}^{\m\n}=B^\m u^\n-B^\n u^\m +\tilde{F}^{\m\n}_{(1)},
\end{eqnarray}
with $\tilde{F}^{\m\n}_{(1)}$ and $B^\m$ transverse to $u^\m$. One can impose the local thermodynamic laws, e.g., the first law and a generalized Gibbs-Duhem relaion, as 
\begin{eqnarray}
	\label{spinmhd}
	& Tds +\frac{1}{2}\m_{\m\n} d S^{\m\n} + H_\m d B^\m= d\ve,&\\
	&Ts+\frac{1}{2} \m_{\m\n} S^{\m\n} + H_\m B^\m=\ve+P,&
\end{eqnarray}
with $H_\m$ the ``magnetic potential" conjugate to the magnetic flux (physically, it can be interpreted as the in-medium magnetic field strength). The convariant form for the Gibbs-Duhem relation is
\begin{eqnarray}
	\label{firstlawmhd}
	s^\m &=& P\b^\m +\H^{\m\n}\b_\n-\frac{1}{2}\S^{\m\r\s}\a_{\r\s} +\tilde{F}^{\m\n}\g_\n,
\end{eqnarray}
with $\g^\m = \b H^\m$. The second law of thermodynamics requires $\pt_\m s^\m\geq 0$, which imposes constraints on the possible forms of the constitutive relations order by order in the gradient expansion. Recently, such a framework for spin magnetohydrodynamics has been discussed; see Refs.~\cite{Fang:2024sym,Fang:2024hxa} for more details. 

It would be very interesting to extend these studies to include possible parity-violating effects, thereby obtaining spin magnetohydrodynamics in a chiral conducting medium. This would provide a bridge between spin magnetohydrodynamics and chiral magnetohydrodynamics. Another issue that may affect the formulation of spin magnetohydrodynamics is the pseudo-gauge ambiguity. As we have seen, such an ambiguity is crucial for the formulation of spin hydrodynamics, and it would be interesting to explore how it influences the formulation of spin magnetohydrodynamics. Finally, exploring possible collective modes and instabilities in such a fluid will also be important. This would be valuable for potential applications (e.g., possible dynamo mechanisms due to spin degrees of freedom) in what we might call spin plasma, whether in heavy-ion collisions or astrophysical systems.

{\bf (2) Calculation of new transport coefficient.} As we have seen, new transport coefficients appear in spin hydrodynamics, most notably the rotational viscosity $\eta_s$. Strictly speaking, $\eta_s$, unlike the usual shear viscosity $\eta$, is not a transport coefficient in the traditional sense. It does not characterize the ability to transport spin within the fluid; rather, it represents how quickly the spin density relaxes to its equilibrium value determined by thermal vorticity. This can be easily understood by rewriting \eq{con:Mmnr2} in the canonical pseudo-gauge and in component form (keeping linear terms in spin density and velocity): $\pt_t S^i\approx-\eta_s (\m^i-\varpi^i)$ where $\m^i=\e^{ijk}\m_{ik}$, which leads to $\pt_t\m^i=-\G_s(\m^i-\varpi^i)$ with $\G_s=\eta_s/\c_s$ the spin relaxation rate and $\c_s$ the spin susceptibility. Nevertheless, the calculation of $\G_s$ and, equivalently, $\eta_s$ is important for understanding the evolution of spin polarization. Recently, $\G_s$ has been computed perturbatively for heavy quarks in hot QCD plasma~\cite{Hongo:2022izs,Li:2019qkf} and for baryons in hot hadronic plasma~\cite{Hidaka:2023oze}. Kinetic theory-based calculations have also been reported. The results show that for heavy quarks, this parameter can be parametrically small, making the spin degree of freedom a quasi-hydrodynamic mode. In the future, the calculation of other new transport coefficients, such as those arising in gyrohydrodynamics~\cite{Cao:2022aku}, could also be crucial for understanding spin dynamics in different fluids. Additionally, it will be important to examine and understand the pseudo-gauge dependence of these new transport coefficients.

{\bf (3) Simulation of spin hydrodynamics.} In order to apply spin hydrodynamics to, for example, heavy-ion collisions, it is important to develop a suitable numerical framework for performing simulations. It is well known that first-order relativistic hydrodynamic equations suffer from numerical instabilities and the emergence of acausal modes. The origin of this problem lies in the fact that first-order constitutive relations are non-dynamical, meaning that the response of the fluid to thermodynamic forces is instantaneous. One solution to this problem is to make the constitutive relations dynamical. For example, the constitutive relation for the shear channel can be modified to
\begin{eqnarray}
	\label{isconst}
\t_\p (D\p)^{\m\n}+\p^{\m\n}=2\eta\s^{\m\n},
\end{eqnarray}
with $\p^{\m\n}$ being the tracelss symmetric part of $\H^{\m\n}_{(1)}$, $(D \p)^{\m\n}\equiv(1/2)[\D^{\m\r}\D^{\n\s}+\D^{\m\s}\D^{\n\r}-(2/3)\D^{\m\n}\D^{\r\s}]D\p_{\r\s}$ being the traceless part of the co-moving time derivative of $\p_{\m\n}$, and $\t_\p$ being introduced to represent how quickly $\p^{\m\n}$ relaxes into the hydrodynamic constitutive relation. (Note that this procedure introduces new dynamic mode that is not hydrodynamic mode, and would relax on a timescale given by $\t_\p$.) The use of such a modification has been very successful in the numerical simulation of relativistic hydrodynamics. For relativistic spin hydrodynamics, to implement numerical simulations, one may adopt similar modifications to the constitutive relations. This has been recently discussed in Refs.~\cite{Liu:2020ymh,Weickgenannt:2022zxs,Weickgenannt:2022qvh,Biswas:2023qsw,Xie:2023gbo,Daher:2024bah,Wagner:2024fry}. Essentially, the constitutive relation (\ref{tmnafirstphi}) is replaced by a dynamic relation 
\begin{eqnarray}
	\label{isconstphi}
\t_\f (D \f)^{\m\n}+\f^{\m\n}=\eta_s\D^{\m\r}\D^{\n\s} \lb \m_{\r\s}-T\varpi_{\r\s}\rb,
\end{eqnarray}
with $\t_\f$ a relaxation time for the antisymmetric part of the energy-momentum tensor and $(D\f)^{\m\n}\equiv\D^{\m\r}\D^{\n\s} D\f_{\r\s}$ being the transverse part of the co-moving time derivative of $\f_{\m\n}$. With these modifications, a numerical simulation of relativistic spin hydrodynamics can be performed, which will provide valuable insights into spin polarization phenomena (see recent progresses in Refs.~\cite{Singh:2024cub,Sapna:2025yss}), such as those observed in heavy-ion collisions. 

\textbf{Acknowledgement ---} We acknowledge the collaborations on topics of spin hydrodynamics with Zheng Cao, Koichi Hattori, Masaru Hongo, Matthias Kaminski, Yu-Chen Liu, Mamoru Matsuo, Yu-Shan Mu, Shi Pu, Misha Stephanov, Hidetoshi Taya, Ho-Ung Yee, Zhong-Hua Zhang, and insightful discussions with Francesco Becattini, Matteo Buzzegoli, Wojciech Florkowski, Kenji Fukushima, Umut Gursoy, Defu Hou, Jin Hu, Jinfeng Liao, Dirk Rischke, Xin-Li Sheng, Enrico Speranza, David Wagner, Qun Wang, and Amos Yarom. This work is supported by the Natural Science Foundation of Shanghai (Grant No. 23JC1400200), the National Natural Science Foundation of China (Grants No. 12225502, No. 12075061, and  No. 12147101), and the National Key Research and Development Program of China (Grant No. 2022YFA1604900).

\bibliography{reference}

\end{document}